\documentclass[aps,pre,twocolumn,preprintnumbers,amsmath,amssymb]{revtex4}
\usepackage{graphicx}
\begin{document}
\title{On the heat capacity of liquids at high temperatures}

\author{S.M. Stishov}
\email{sergei@hppi.troitsk.ru}
\affiliation{Institute for High Pressure Physics of RAS, 108840, Troitsk, Russia}

\begin{abstract}
Making use of a simple approximation for the evolution of the radial distribution function, we calculate the temperature dependence of the heat capacity $C_v$ of Ar at constant density. $C_v$ decreases with temperature roughly according to the law $\sim T^{-1/4}$, slowly approaching the hard sphere asymptotic value $C_v=\frac{3}{2}R$. However, the asymptotic value of $C_v$ is not reachable at reasonable temperatures , but stays close to 1.7--1.8 $R$ over a wide range of temperatures after passing a " magic " $2R$ value at about 2000 K.  Nevertheless these values has nothing to do with loss of vibrational degrees of freedom, but arises as a result of a temperature variation of the collision diameter $\sigma$.
\end{abstract}
\maketitle

It has been known for a long time that $C_v$ of many liquids almost equals that of solids at the melting point, decreases with temperature, and reaches a value of $\sim 2 R$ ($R$-gas constant)~\cite{1,2,3,4,5,6}. A similar trend was observed in model systems of particles interacting with an inverse power potential~\cite{7}. With time, a general belief was formed that $C_v$ varies from $\sim 3R$ to $\sim 2R$ with temperature from the melting to the critical temperature. Ref.~\cite{8} provides an attractive, though probably wrong, explanation of this behaviour of $C_v$. It assumes that, in liquids at high temperatures, the transverse vibrational modes cease to exist, whereas the longitudinal mode still remains, resulting in $C_v = 1R+1/2R+1/2R=2R$.  Recently, it has been proposed that decreasing the heat capacity beyond the $2R$ value implies a transition to a truly gaseous state without any vibrational degrees of freedom. A corresponding line in the $P$-$T$ space bordering a region with $C_v\geq2R$ was named a Frenkel line in honor of the Russian theoretician Yaakov Frenkel~\cite{9,10,11}.

However, the interpretation of the $2R$ value of a heat capacity as a result of loss of the transverse elastic response is not proved in either case. So it is appropriate to carry out an analysis of the situation based on a traditional approach to the thermodynamics of liquids and gases.

Let’s write the energy of a classical system of particles in the form,
\begin{equation} \label{eq1}
E=\frac{3}{2}RT+\frac{N^2}{2V}\int\limits_{0}^{\infty}\Phi(r)g(r)4\pi r^2dr
\end{equation}
where   $\Phi(r)$ – pair interaction potential, $g(r)$ - radial distribution function, $N$ - Avogadro number,  $R$ - gas constant.

We approximate the radial distribution function $g(r)$ at moderate density by the step function,
\begin{eqnarray} \label{eq2}
g(r)=0, at\ r<\sigma\\ \nonumber
g(r)=1, at\ r>\sigma
\end{eqnarray} 

This approximation should work well enough at high temperatures.
Then the expression~\ref{eq1} takes the form,
\begin{equation} \label{eq3}
E=\frac{3}{2}RT+\frac{N^2}{2V}\int\limits_{\sigma}^{\infty}\Phi(r)4\pi r^2dr
\end{equation}

For further analysis we will use for $\Phi(r)$ the Lennard-Jones~\cite{6,7,8,9,10,11,12} potential,
\begin{equation} \label{eq4}
\Phi(r)=4\epsilon\left[ \left( \frac{r_0}{r}\right) ^{12}-\left( \frac{r_0}{r}\right) ^6\right] 
\end{equation}
After substitution eq.~\ref{eq4} in eq.~\ref{eq3} and integration one obtains,
\begin{equation} \label{eq5}
E=\frac{3}{2}RT+\frac{8\pi\epsilon N^2}{V} \left[ \frac{1}{9}\frac{{r_0}^{12}}{\sigma^9}-\frac{1}{3}\frac{{r_0}^6}{\sigma^3}\right] 
\end{equation}
Now follows the important step in this approach, assuming a form of  the evolution $g(r)$ of $\sigma$ with temperature. In the framework of the $g(r)$ definition~\ref{eq2} we can write the temperature dependence of $\sigma(T)$ using,
\begin{equation} \label{eq6}
\epsilon \left( \frac{r_0}{r}\right) ^{12}=\frac{1}{2}kT
\end{equation}
where $r=\sigma$

\begin{figure}[htb]
\includegraphics[width=80mm]{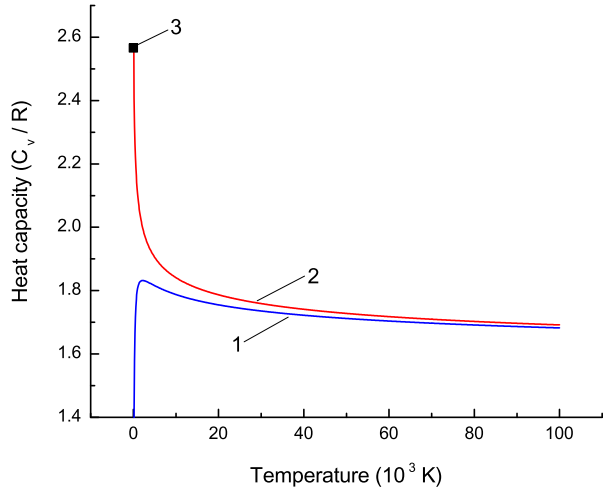}
\caption{\label{fig1} (Color online) Heat capacity $C_v$ of the virtual Ar at a constant volume equals to a value at the triple point. Calculations were made with use the full potential 1, as well as with only repulsive part of it 2, $C_{v}$ of Ar in the triple point 3 Ref.~\cite{13}.}
\end{figure}

To simplify the procedure, we use only the repulsive branch of the potential~\ref{eq4}, though it could influence the low temperature behavior of the following calculation. Finally for the heat capacity $C_v$ we have,
\begin{equation} \label{eq7}
\frac{C_v}{R}=\frac{3}{2}+\frac{\pi {r_0}^3}{3V}N\left[ \left(\frac{2\epsilon}{kT}\right)^{1/4}-\left(\frac{2\epsilon}{kT}\right)^{3/4} \right]
\end{equation}
As is seen from eq.~\ref{eq7}, the heat capacity $C_v$ at constant density of a system of particles interacting with the Lennard-Jones potential decreases at high temperatures roughly according to the law $\sim T^{-1/4}$,  slowly approaching the hard sphere asymptotic value of $C_v=\frac{3}{2}R$. Using the potential parameters known for Ar and the molar volume value of Ar at the triple point ($\varepsilon/k=119.3$, $r_0(A)=3.405$, $V_{T.P.}=28.33$)~\cite{12} and eq.~\ref{eq7} we made the corresponding calculations, shown in Fig.~\ref{fig1}. As we expected our approach does not work properly with the full potential~\ref{eq4} at low temperatures, whereas using only the repulsive part of the potential~\ref{eq4} gives better results. The latter is not surprising  because a system of soft spheres on a homogeneous attractive background works satisfactory for noble gases. Note that the homogeneous attractive background does not contribute to the heat capacity. Remarkably, the reference data for $C_{v}$ in the triple point 
(see Ref.~\cite{13}) lies almost exactly on the calculated curve 2 (Fig.~\ref{fig1}). 
As is seen in Fig.~\ref{fig1} the asymptotic value of $\frac{3}{2}R$ is not easily achieved. The significant part of the heat capacity curves are situated close to the value of $1.8R$, passing a "magic" value of $2R$ at about 2000 K. Even at $10^{5}$ K (compared to the temperature of the Ar critical point 150 K) the heat capacity is still very close to $1.7R$. However, this value is not a result of loss of the vibrational modes, but it occurs in our model owing to the temperature variation of the collision diameter $\sigma$.

In conclusion, making use a simple approximation for an evolution of the radial distribution function, we calculate the temperature dependence of heat capacity $C_v$ of Ar at constant density. The result shows that $C_v$ of Ar decreases with temperature according to the law $\sim T^{-1/4}$ and tends to the hard sphere asymptotic value $C_v=\frac{3}{2}R$. However, as is seen in Fig.~\ref{fig1}, the asymptotic value of $C_v$ is not reachable at reasonable temperatures, and $C_{v}$ stays close to 1.7--1.8 $R$ over a wide range of temperatures.  Nevertheless these values as well as the " magic " value  of $2R$ have nothing to do with loss of the vibrational degrees of freedom, but arise as a collision effect.
\section{Acknowledgements}
This work was supported by the Russian Foundation for Basic Research (grant 15-02-02040), Program of the Physics Department of RAS on Strongly Correlated Electron Systems and Program of the Presidium of RAS on Strongly Compressed Matter. The assistance of Alla Petrova and Marc Costantino is highly appreciated.

\end{document}